**Preclinical Water-Mediated Ultrasound Platform using Clinical FOV for Molecular Targeted Contrast-Enhanced Ultrasound**

Stavros Melemenidis[1], Anna Stephanie Kim[2], Jenny Vo-Phamhi[1], Edward Graves[3], Ahmed Nagy El Kaffas[3]*, Dimitre Hristov[1]*.

1. Department of Radiation Oncology, Stanford University School of Medicine, Stanford, CA 94305, USA.
2. Tuft University School of Dental Medicine, 1 Kneeland St, Boston, MA, 02111, USA.
3. Department of Radiology, University of California, San Diego, La Jolla, CA 92037, USA.

**Corresponding equal contribution senior authors:**

Dimitre Hristov, PhD

*Department of Radiation Oncology,*

*Division of Radiation Physics,*

*Stanford University School of Medicine,*

*Stanford, 94305 CA, USA*

*Email:* [dhristov@stanford.edu](mailto:dhristov@stanford.edu)

Ahmed Nagy El Kaffas, PhD

*Department of Radiology*

*University of California, San Diego,*

*La Jolla, 92037 CA, USA*

*Email:* [aelkaffas@health.ucsd.edu](mailto:aelkaffas@health.ucsd.edu)

**Disclosures:**

The authors have nothing to disclose.

**ABSTRACT**

**Background:** This protocol introduces an ultrasound (US) configuration for whole-body 3D dynamic contrast-enhanced ultrasound (DCE-US) imaging in preclinical applications. The set-up relies on a clinical abdominal matrix US probe to enable mice imaging beyond current preclinical systems that are generally unable to capture whole-body volumetric and dynamic imaging. We demonstrate via this protocol the feasibility of volumetric contrast-enhanced and molecular 3D imaging in the entire lower body of mouse, as well as the capability of imaging multiple lesions in the same animal simultaneously with a single contrast bolus injection.

**Methods:** We modified a silicone cup with a 101.6 mm inner diameter and 6.4 mm wall thickness, to a height of 76.2 mm and cut out a rectangular side window (12.7 × 15.9 $mm^2$) at the lower part of the cup. Mice were positioned with their head and the two front legs outside of the cup while the rest of the body remained inside the cup and submerged under water; the edges around the mouse's body in contact with the cup's wall were sealed with Vaseline to prevent water leakage, and the cup was filled with warm water to maintain the mouse's body heat. Imaging was conducted using a portable clinical ultrasound system (EPIQ 7 Philips) with an abdominal matrix-array transducer (PM mode, X6–1) inserted into the top of the cup, and positioned to visualize the entire of the mouses abdomen. The mice were imaged before and after the injection of P-selectin-targeted microbubbles.

**Results:** Here, in nu/nu mice with two adjacent tumors, we demonstrate that a commercially available clinical matrix transducer can be utilized to achieve whole-body 3D DCE-US and molecular US (mUS) imaging, as well as capture independent qualitative and quantitative information from several lesions or organs at different locations in the animal, through a single bolus injection.

**Conclusion:** Our set-up is simple, inexpensive and readily available if users have access to a matrix probe with contrast imaging capabilities. It allows for standardized comparison of perfusion properties within the same animal of different lesions and/or organs across the whole of the animal body.

**INTRODUCTION**

Ultrasound (US) is a non-invasive and cost-effective modality for pre-clinical imaging in small animals [1–5]. Mice are the most used species in pre-clinical research, and US is suitable for anatomical and functional imaging of all organs in the mouse body with exception of the lungs and head (although both can be overcome). In contrast mode, also known as dynamic contrast-enhanced ultrasound (DCE-US), the diagnostic potential of ultrasound is significantly enhanced through the use of microbubbles, which resonate in response to ultrasound waves due to their pronounced acoustic impedance mismatch with surrounding tissues, enabling real-time visualization of blood flow and vascular structures. The development of highly echogenic ultrasound contrast microbubbles has drastically expanded the diagnostic potential of ultrasound, enabling exclusive imaging of several dynamic tissue perfusion properties and vascular molecular imaging. Most importantly, microbubbles have non-linear properties that enable high resolution/sensitivity and SNR that can enable the detection of a single microbubbles [6,7]. Nonetheless, in the context of pre-clinical imaging, contrast-enhanced ultrasound remains limited to 2D imaging for assessing dynamic perfusion, with small fields of view that restrict the investigation to a single organ per acquisition. This is because current transducers tailored for small animal imaging employ high center frequencies to maximize spatial resolution at the cost of penetration depth and are not designed for dynamic 3D imaging. Volumetric imaging can only be obtained via stepwise sweep acquisitions of 2D serial images over a z-distance, which makes the volume non-dynamic. For example, Fujifilm has a well-developed a preclinical 3D-Mode where a distance in Z direction can be set and robotically scan the area, but with small field size probes and not does not provide 3D in real time. Another innovation for 3D US imaging is the accessory Piur-Tus-infinity (Piur-Imaging) that tracks probe's movement when attached to ultrasound transducer and provides tomographic reconstruction, but with not real time capability. Several commercially available matrix probes have been developed in the clinic for 3D imaging. In the context of mice imaging, these could offer dynamic whole-body imaging given their footprint, and despite their low-resolution, our group has extensively demonstrated the potential use of such clinical probes in mice when used with microbubble contrast agents [8–11].

We present here a protocol using an *in vivo* platform for volumetric whole-body (excluding the head) contrast mode US imaging of small animals and suitable for probes with wide transducer matrices. Our platform is a water bath for dissociative/sedative or generic anesthesia anesthetized mice (ketamine/xylazine or isoflurane respectively) where a stationary probe can be placed at a distance from the mouse and provide volumetric imaging of the entire abdomen, dynamically tracking contrast signals throughout the mouse's abdomen. In contrast mode, it allows for dynamic simultaneous assessment of perfusion in several organs or tissues following a single bolus, which provides systemic functional information based on both microbubble flow, suitable also for other forms of microbubbles. Targeted microbubbles add the potential to assess molecular expression of specific markers on endothelial cells in multiple organs and determine clearance. Capturing both anatomical and functional data, the system can be used internal control studies with various pre-clinical diagnostic and pharmacological interests. Our platform aims to showcase that with no additional novelty, existing probes technology and previously explored ideas of water as media, we can utilize both contrast and molecular US information, offering comprehensive insights into perfusion dynamics throughout the mouse body. We showcase our set-up in an experimental tumor mouse model with two contralateral tumors (double tumor model), demonstrating that both tumors can be assessed simultaneously with contrast and molecular ultrasound. Double tumor mouse models facilitate the investigation of treatment strategies (e.g., drug combinations), abscopal effects, and metastases [12,13].

**EXPERIMENTAL SECTION**

All animal experiments and procedures were approved by the Institutional Animal Care and Use Committee of Stanford University (APLAC-17186) in accordance with institutional and NIH guidelines. All mice anesthesia was induced with 3% isoflurane in $O_2$ at a flow rate of 2 L/min and was maintained with 1.5% isoflurane in $O_2$ at a flow rate of 2 L/min.

The configuration used for the experiment requires two working stations; Station A where a tail IV catheter is secured in the mice under anesthesia, and Station B where the mice are positioned in the water-mediated US platform for imaging under anesthesia (**Figure 1A**).

1. Station A - Secure tail IV catheter under anesthesia

Station A comprises from an anesthesia nose cone (isoflurane vaporizer) and a heating blanket set at 36°C to avoid loss of mouse body heat when under anesthesia (**Figure 1A**; Station A). For IV catheter, a butterfly needle with convenient gauge (1 per mouse), a compatible silicon tubing and 1 mL syringe (2 per mouse) was used. The dead volume of the catheter's tubing was measured using injectable saline, and after the catheter's insertion we confirmed that there is no resistance upon injectable volume. To secure the catheter on the tail, we use labeling tape (Rainbow, Genemate) to fix in place the butterfly needle to the tail and size down tooth peak to keep the tail from bending. We cover the entire plastic area of the butterfly needle, the tail and the tooth peak with tape from both sides, and tape them carefully together to maximize sealing and minimize water from emerging at the needle. After confirmation of catheter's functionality the excess tape beyond the plastic area of the butterfly needle was removed with scissors and confirm that there is no sticking part exposed (**Figure 1B**; red).

2. Station B - Water-mediated US platform

Station B comprises of a tray, a secondary anesthesia nose cone, a secondary heating blanket set at 36°C, a silicone cup and the US probe (**Figure 1A**; Station B). We used a silicone measuring cup with a 101.6 mm inner diameter and 6.35 mm wall thickness, and modified it by reducing its height to 76.2 mm. Additionally, a rectangular side window (12.7 × 15.9 $mm^2$) was cut out at the lower part of the cup allowing for the positioning of a mouse from inside the cup, by passing the head and the two front legs outside the cup while the rest of the body remains inside the cup. After the positioning the mouse, the mouse head is placed at the anesthesia nose cone (**Figure 1B,C**). To prevent any water leakage from air gaps between the mouse's body and the walls of the cup's rectangular window, Petroleum Vaseline was used for water-proof sealing. We used a tray that can hold water in case there is a leak, placed on the top of a heating blanket that maintained surface temperature of 36°C. In order to hold the probe is place, we used a table mount flexible gooseneck clamp at the end of the working table.

3. In Vivo Working protocol

After securing the IV catheter to the anesthetized mice, remove mouse and IV catheter from Station A, move to Station B and place the mouse directly at the anesthesia nose cone. Apply Vaseline at the front legs and neck of the mouse. Move the mouse inside the silicon cup and leave the saline syringe outside the cup. Place the mouse's nose at the opening of the cup and move the silicon cup's opening next to the anesthesia nose cone to maintain anesthesia. Gently pull the head of the mouse outside of the silicone cup's opening, until it can rest comfortably in the nose cone. Use plastic tweezers to gently pull the two front legs, one at the time, out of the opening, while the mouse is breathing through the nose cone. Only the head and the two legs should be outside the opening while the rest of the mouse should be spread out flat in the cup with the tumors close to the center of the cup. Elevate the head of the mouse and the nose cone, such as in the case of a water leakage the mouse will be safe (**Figure 1B,C**). Apply only the necessary amount of Vaseline inside and outside the cup's opening window to waterproof the opening. Don't be excessive with Vaseline to avoid contaminating the water. Prepare water at 37°C and confirm the temperature with a thermometer in the beaker. Slowly pour 400 mL of water into the platform where the mouse is placed. Avoid turbulence and bubbles in the water. Remove any bubbles from the body using plastic tweezers. Place a waterproof object onto the tail that is heavy enough to minimize respiration motion and buoyancy of the mouse body. **Figure 1B** shows an example were a magnetic stirrer was used. Test whether the tail line is functional by perfusing a small volume or withdrawing a volume until blood is observed. Place the probe into the water and find the desired FOV on the mouse.

4. In Vivo US imaging Protocol

Images were acquired with a clinical portable US system (EPIQ 7 Philips) coupled to an abdominal matrix-array transducer (PM mode, X6–1; Philips US, Bothell Wash). Note that other probes could be used here, and Philips does offer other matrix probes that may be more suitable depending on the application (i.e. X14-1, X5-1). The probe was positioned by localizing the tumors in 2D B-mode imaging. The positioning was fine-tuned in 3D and then fixated in place for the rest of the imaging. Several parameters were optimized uniquely for our longitudinal imaging study; these should be set by the user for optimal imaging. In this protocol, we used: persistence OFF, Flash frames set to 3, and focus set to 5 cm. In our proof of principal example, mice were imaged 24 hours after radiation therapy of one of the adjacent tumors, which was implemented to demonstrate specific uptake of targeted microbubbles measuring Differential Targeted Enhancement (DTE) of both tumors. Rather than continuing acquisition across the whole imaging session, the contrast mode 3D acquisition occurred for 30 seconds before and 2 minutes immediately post IV of targeted microbubbles, 30 seconds at 5 minutes post IV, and 30 seconds before and 2 minutes post bursting pulse (**Figure 1D**); imaging acquisition schedule should be customized per user needs. Quantification of the DCE-US signal was carried out by measuring DTE. The imaging sequence and workflow should be optimized by the user accordingly.

5. US Targeted Microbubble Preparation

VEVO MicroMarker microbubbles (FUJIFILM | Visual Sonics, Toronto, ON, Canada) were used for the development of our platform; any microbubbles or targeted microbubbles can be used for this protocol. The microbubbles were reconstituted with 1mL of saline according to the manufacturer protocol. For p-selectin targeted microbubbles, rabbit anti-mouse CD62P primary antibodies (ab202983; Abcam, USA) were added to the reconstitute according to the guidelines

of the manufacturer. From the final product, 100 µL volume per animal per imaging session was administered via tail catheter within a period of maximum two hours after preparation.

6. Image Analysis

Image analysis can be carried out in any software that allows for volumetric imaging. In our case, we used MevisLab and developed custom tools to visualize and draw volumes of interest VOIs. Other software options we have used include ITKsnap [14]. Custom software tools can be made available to potential end users upon request.

7. Animal Model – Double Tumor

6-8 weeks old female Nu/Nu mice were purchased from Jackson Laboratory (CA, USA) and given a 2-week acclamation period. Each mouse under anesthesia was inoculated with two subcutaneous injections (100 µL each) of $5 \times 10^5$ Lewis Lung Carcinoma [LL/2/(LLC1] cells, in each of the upper hind limbs. When tumors reached volume of 120-180 mm$^3$, ensuring presence of tumor vasculature, one of the two tumors was irradiated with a single fraction of 6 Gy.

8. Tissue Culture and Tumor Cell Injection Preparation

Work aliquots were thawed in 37 ºC water bath and seeded in vented tissue culture treated flasks with DMEM medium containing 10% FBS. After two passages (splitting) the cells reached the required number and prepared for injections. Injections were prepared in a final concentration of $5 \times 10^6$/mL (100 µL/injection).

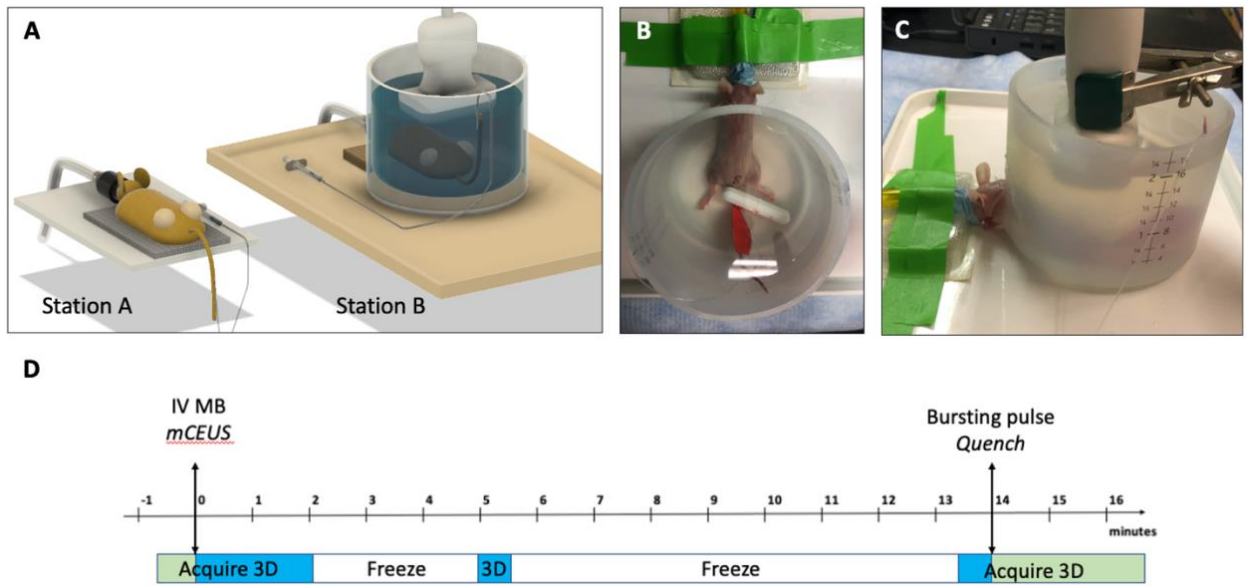

**Figure 1**

US water-mediated platform setup and imaging protocol. A) Computer-aided design (CAD) of the in-vivo imaging experimental set-up. Station A is used for the preparation of waterproof IV catheters on the anesthetized mouse and subsequently is transferred to the imaging Station B. B) Top view photograph of the imaging platform, illustrating the placement of mice inside the platform after being filled with water. At the tail the red tape ensures the IV catheter in place and a stirring magnet (white) is used to minimize respiration motion of the mouse's body. C) Side view photograph of the platform showing the placement of the probe, which is suspended by flexible gooseneck clamp. D) Schematic of the imaging protocol of a single microbubble bolus; 30 seconds before and 2 minutes immediately after the IV of targeted microbubbles, 30 seconds at 5 minute post IV, 30 seconds before and 2 minutes immediately after bursting pulse (total acquisition time of 5:30 minutes).

**RESULTS**

We have been able to replicate and reproduce the protocol consistently, demonstrating the feasibility of the configuration. Mice have been under anesthesia for a maximum of 30 minutes and under water for a maximum of 20 minutes. Mice recovered from anesthesia within 3 minutes and regained vitality within 10 minutes. Overall, no animal wellness complications or mortalities were observed associated with the procedure.

From the anatomical US (B-mode) dataset, volumetric images of the lower half of the mice were reconstructed in MevisLab, showing both adjacent tumors (**Figure 2A-H**). Tumor segmentation of parallel axial sections allowed tumor volume measurements (**Figure 2A-H**; blue and purple). The right and the left kidney were distinguished.

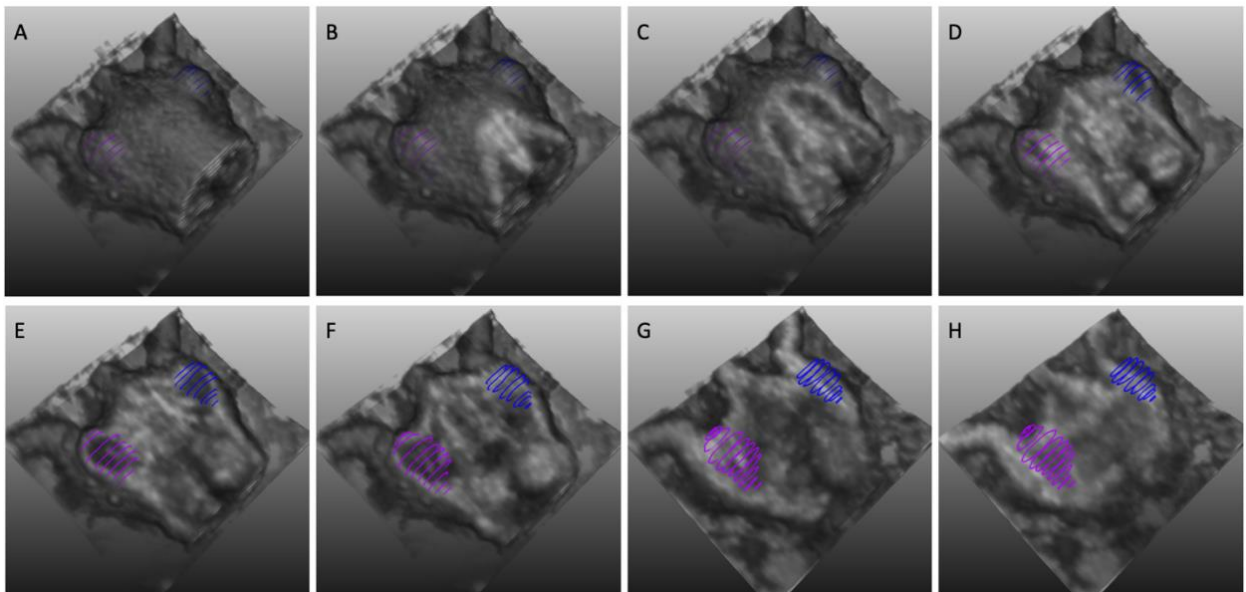

**Figure 2**

US A-mode 3D dataset reconstruction of a double tumor model, illustrated through different coronal planes. The FOV is set bellow the liver. The adjacent subcutaneous tumors are segmented at representative axial planes across the tumor (left tumor in blue and right in magenta).

In our setup, we have included both hind-leg tumors on both sides within the imaging FOV for concurrent imaging of contrast. In pre-clinical investigations where tumors are implanted higher up in the body (or absent in non-cancer studies), the FOV can be set higher to include the liver of the mouse and provide a whole-abdomen imaging dataset.

Using the dynamic contrast mode dataset, we can reconstruct 3D images that allow for observation of functional information, like the uptake and wash out of the contrast agent over time. **Figure 3A-H** shows a representative coronal height of the 3D reconstruction of the same animal as **Figure 2** at different time points post IV contrast agent. The baseline of the contrast mode was acquired for 30 seconds prior to (**Figure 3A**) and 2 minutes after (**Figure 3B**) IV of the contrast agent. In **Figure 3B**, tumors show rapid uptake and reach the maximum contrast. Thereafter, the contrast was reduced over time (**Figure 3C-D**) until microbubbles were quenched and intensity returned to baseline (**Figure 3E**). Tumors show good perfusion in their periphery but not in their center, which is consistent with histological staining (data not shown).

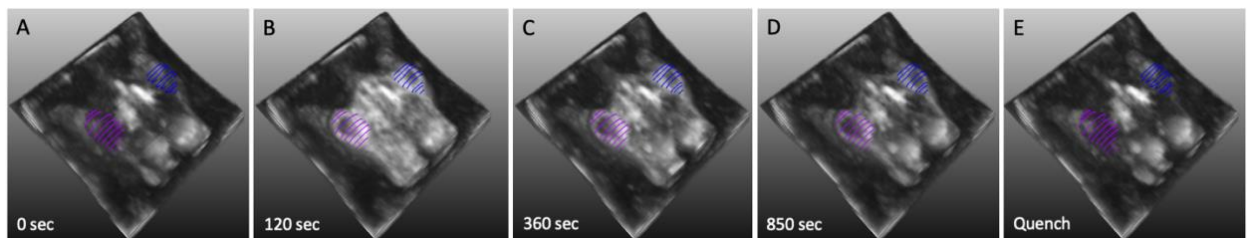

**Figure 3**

B-mode 3D DCE-US in vivo imaging dataset reconstruction of a double tumor model, illustrated at a fixed coronal plane and at different time points before and after injection of molecular targeted MB. A) Baseline image contrast prior to MB injection. B) Contrast peak at 2 minutes post MB injection. C-D) Contrast decay over time. E) Contrast baseline after MB bursting.

Time intensity curves (TICs) of the adjacent tumor ROIs are shown in **Figure 4A**. The irradiated tumor shows higher contrast agent uptake (right tumor; magenta color), which indicates higher expression of p-selectin 24h hours after 6 Gy irradiation. Representative B-mode with contrast mode of the acquisition points shows hot spots of the contrast agent in the tumor and the kidneys.

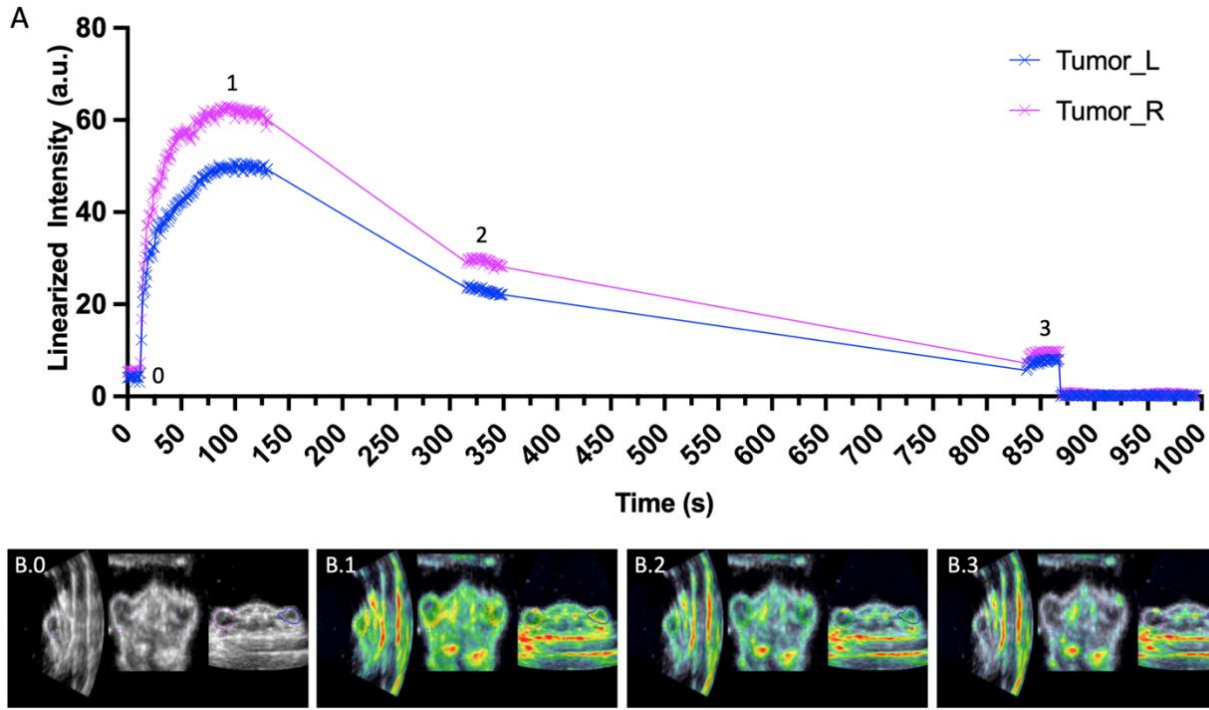

**Figure 4**

Differential targeted enhancement of adjacent tumors and 3-plane B-mode DCE-US. A) Time intensity curve (TIC) of the contrast uptake at the volumes of two adjacent tumors, 24 hours post irradiation of 6Gy of the right tumor. The contrast mode acquisition was discontinued between the indicated time points (1-3) to save data space. Microbubbles were burst (quenched) after 14 minutes. B.0-3) 3-plane B-mode DCE-US across the first 14 minutes, representing points in imaging duration 0-3 as indicated on (A).

## DISCUSSION

In this proof-of-concept work, we present a platform for volumetric, molecular targeted 3D DCE-US for abdominal imaging of mice. Using water as a coupling medium, which minimizes air bubble artifacts that often arise with gel-based imaging and are significantly more pronounced in pre-clinical imaging, we increased the distance of the probe from the mouse surface. Consequently, we have captured a wider FOV and demonstrated partial abdominal US volumetric imaging that provides anatomical information for multiple organs in a single acquisition. By using a double tumor mouse model, we accomplished volume and simultaneous dynamic blood perfusion measurements of both tumors with a single anatomical volumetric image, as well as molecular characterization of both tumor volumes using the same contrast injection (bolus); this enables direct one-to-one relative quantification given that both tumors receive the same dose of contrast agent. Such multi-anatomical information for multiple organs is not available when imaging tumors with a conventional US system given the small FOV and the lack of real 3D acquisition.

The concept of using water as coupling medium is an old one and it has been recently demonstrated in an experimental setup of ultrafast Doppler ultrasound imaging of the murine kidney [15]. However, in this set up a high resolution probe was utilized and wide field real 3D acquisition was not demonstrated. We aimed to showcase the implementation of a low resolution sophisticated clinical probes for wide field 3D DCE-US imaging. However, as clinical and preclinical probes advance, the capabilities of the concept introduced here can be modified to utilize any 3D imaging probe; either matrix or row-column. Clinical probes currently range in frequency from 1–10 MHz and intraoperative probes can achieve 20 MHz; the latter remain limited to spatial resolutions in the 100s of microns, which can be less effective for certain animal studies. Additionally, clinical scanners designed to image the human heart with 60–100 beats per minute (bpm) have insufficient temporal resolution to image the rapid heart movement of preclinical models (400–600 bpm). Pre-clinical probes can achieve 10–70 MHz with spatial resolutions capable of resolving structures smaller than 100 microns, and pre-clinical probes are designed with sufficient temporal resolution to resolve cardiac motion within a mouse heart. The Philips XL14-3 achieves even higher frequencies and greater resolutions with a broadband range of 14-3 MHz but does not yet offer contrast mode. 3D DCE-US is a powerful and promising imaging method for characterizing and monitoring lesions with reduced sampling errors, as shown by recent results in humans [16].

The results presented here show that our platform using larger more sophisticated clinical probe with wider 3D arrays has the potential to perform multi-organ diagnostic and theragnostic 3D DCE-US imaging in small animals. The in-house design of our platform is a prototype. Those seeking to replicate our work should carefully note the following: 1) The opening of the cup matches the size of the mouse's chest at a certain weight, so a mass variation of greater than 2–3g can be significant. When the match is good, a minimum quantity of Vaseline can be very effective. In lighter mice, the setup needs extra attention. Furthermore, a heavier mouse might have problems breathing, so a different size opening must be used. 2) In the event of a leakage, the mouse must be protected, and elevation of the mouse's head is essential. We have not tried using mice with fur, but we believe that this would not be a problem for sealing the cup following shaving or removal of hair using hair removal cream products; this is especially important to minimize impact on image quality and reduce US artifacts. 3) The resolution of our system was limited to ~0.5 mm, but this was sufficient for our results due to the increased signal contrast and SNR when using microbubbles as a contrast agent. Our probe was dedicated to research, and we have been able to use it for our study, but these probes are generally less readily accessible and can be expensive to purchase, although relatively more affordable than current major pre-clinical dedicated systems.

Our proof-of-concept demonstrates that the development of commercially available platforms for wide-field preclinical imaging would be of great benefit to biomedical research. Such tools would significantly advance research in domains from the treatment of vascular, gastrointestinal, and systemic disease to pharmacokinetics. The possibilities further multiply when molecular markers are used. We hope this platform will not only aid researchers but also inspire companies to offer lower-power matrix (i.e. 3D imaging) probes with higher FOVs for preclinical investigation using whole-abdominal dynamic imaging.